\documentclass{Interspeech2024}
\usepackage{amsmath,graphicx,epsfig,stfloats,hyperref,multirow,booktabs,caption,color,cite,amssymb,subfig}

\interspeechcameraready

\title{SimuSOE: A Simulated Snoring Dataset for Obstructive Sleep Apnea-Hypopnea Syndrome Evaluation during Wakefulness}

\name{Jie Lin$^{1}$, Xiuping Yang$^{2,\dagger}$, Li Xiao$^{1}$, Xinhong Li$^{1}$,\\ Weiyan Yi$^{1}$, Yuhong Yang$^{1}$, Weiping Tu$^{1,\star}$, Xiong Chen$^{2,\star}$}

\address{
  $^1$NERCMS, School of Computer Science, Wuhan University \\
  $^2$Sleep Medicine Centre, Zhongnan Hospital of Wuhan University}
\email{tuweiping@whu.edu.cn; zn\_chenxiong@whu.edu.cn}


\newcommand\blfootnote[1]{%
\begingroup
\renewcommand\thefootnote{}\footnote{#1}%
\addtocounter{footnote}{-1}%
\endgroup
}

\begin{document}

\maketitle
\begin{abstract}
     Obstructive Sleep Apnea-Hypopnea Syndrome (OSAHS) is a prevalent chronic breathing disorder caused by upper airway obstruction. Previous studies advanced OSAHS evaluation through machine learning-based systems trained on sleep snoring or speech signal datasets. However, constructing datasets for training a precise and rapid OSAHS evaluation system poses a challenge, since 1) it is time-consuming to collect sleep snores and 2) the speech signal is limited in reflecting upper airway obstruction. In this paper, we propose a new snoring dataset for OSAHS evaluation, named SimuSOE, in which a novel and time-effective snoring collection method is introduced for tackling the above problems. In particular, we adopt simulated snoring which is a type of snore intentionally emitted by patients to replace natural snoring. Experimental results indicate that the simulated snoring signal during wakefulness can serve as an effective feature in OSAHS preliminary screening.
\end{abstract}

\section{Introduction}
Obstructive Sleep Apnea-Hypopnea Syndrome (OSAHS) is a prevalent sleep disorder characterized by recurrent upper airway obstruction or collapse during sleep~\cite{strollo1996obstructive}, which is clinically diagnosed using polysomnography (PSG)~\cite{punjabi2008epidemiology,dempsey2010pathophysiology}. While effective, PSG is not universally accessible due to the requirement for expert supervision and analysis, which introduces limitations related to time and cost constraints. However, OSAHS affects 9-24\% of individuals of all ages, with an alarming 90\% remaining undiagnosed~\cite{singh2013proportion}.
Therefore, a number of Machine Learning (ML) methods have sought to perform OSAHS evaluation in a time-efficient and cost-effective way.\blfootnote{$^\dagger$ Equal contribution. $^\star$ Corresponding author. }

Previous studies show that OSAHS is strongly associated with anatomical and functional abnormalities of the upper airway~\cite{fox1989speech}, which leads to some acoustic peculiarities in audio signals of people with OSAHS compared to those of people without OSAHS. 
Based on this conclusion, various approaches have explored methods based on audio signals for initial self-screening for OSAHS.
As one of the most prominent symptoms of OSAHS, \textbf{1) sleep snoring} has been utilized in evaluation methodologies to reflect upper airway obstruction~\cite{fiz2010continuous,mesquita2012all,dafna2013osa,ben2012obstructive,xie2023assessment}.
These methods use snoring sounds in overnight sleep audio to classify subjects in OSA severity groups or to estimate the apnea-hypopnea index (AHI)~\cite{berry2020aasm}.  
For example, Xie et al.~\cite{xie2023assessment} combined age, BMI and acoustic features and used XGBoost regression to predict AHI values and identify OSAHS, achieving an OSAHS diagnosis accuracy of 86.6\%.
However, utilizing sleep snoring to diagnose OSAHS requires high-quality overnight sleep audio and preprocessing to extract the snoring. This makes the OSAHS pre-screening process complex and inefficient.
To address this problem, some studies have attempted to analyze OSAHS using \textbf{2) speech signals}, which try to capture and extract acoustic features from sustained vowels, nasals, or sentences~\cite{fernandez2009assessment,goldshtein2010automatic,kriboy2014detection,simply2019diagnosis,botelho2019speech,ding2021severity,pang2022obstructive,yilmaz2023obstructive,zhang23b_interspeech}. 
Compared to sleep snoring, speech signals can easily be recorded while the patient is awake. 
For example, Ding et al.~\cite{ding2021severity} analyze sustained vowels and nasals and achieve the best accuracy of 78.8\% by extracting LPCC audio features.
However, speech signals are emitted by vocal cord vibration or produced in the laryngeal and supra-laryngeal parts~\cite{pevernagie2010acoustics}, which encompasses some but not all information related to upper airway obstruction. This partial coverage could potentially adversely impact screening performance.

\begin{figure}[t] 
  \centering
  \includegraphics[width=0.85\columnwidth]{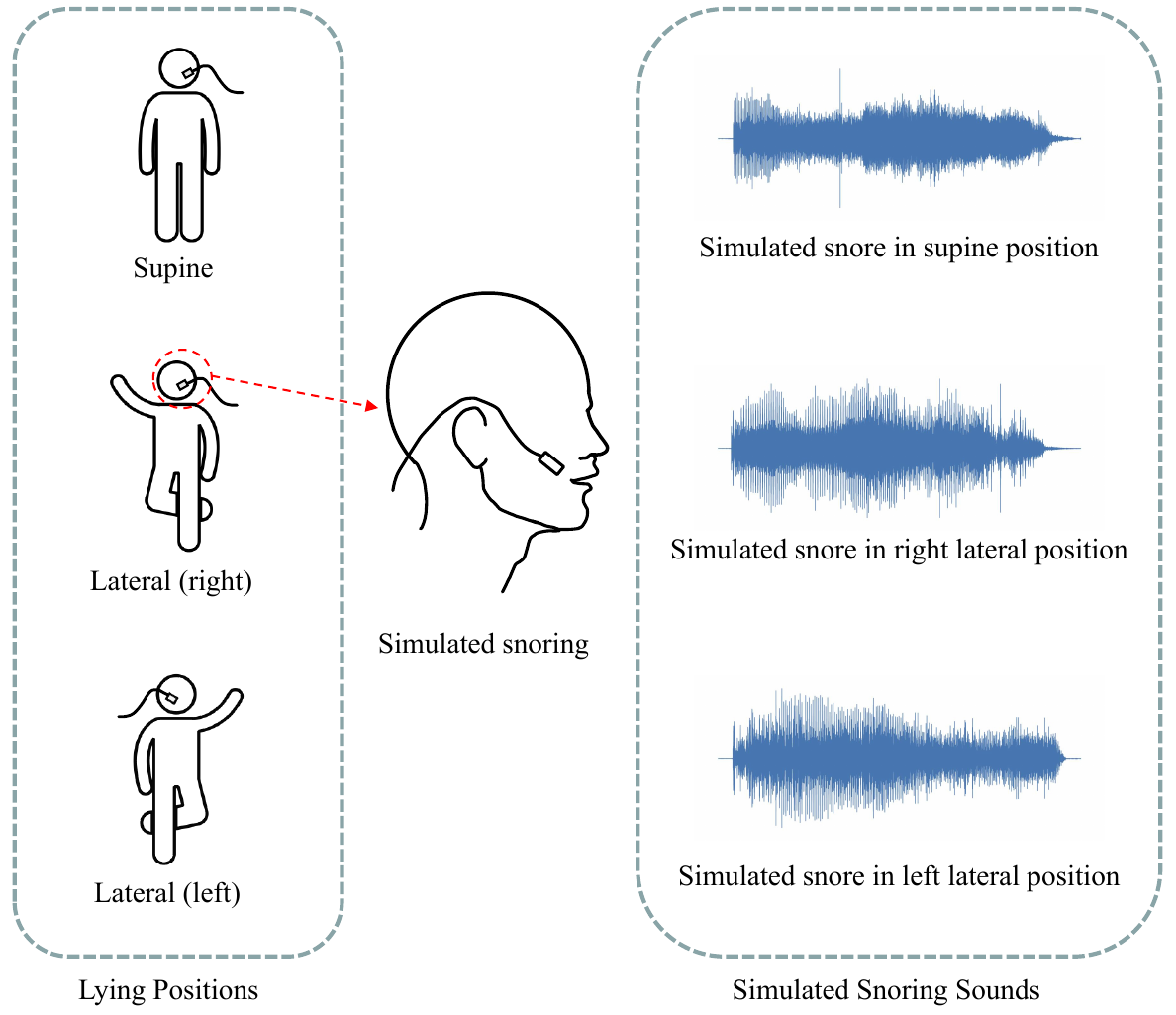}
  \caption{Recording process for simulated snoring. We record snoring sounds in both supine and lateral positions. When recording the subject's snoring in the lateral position, their entire torso faces to the corresponding side. The microphone is secured 3 cm from the mouth on the patient's face.}
  \label{fi:record}
\end{figure}

Research has demonstrated the potential of audio signals for evaluating OSAHS with reasonable performance. Nevertheless, there are still some disadvantages that need to be addressed:
\textbf{1) The collection of sleep snoring is time-consuming.}
Requiring a quiet environment as well as all night, the process of recording sleep audio is time-consuming and inconvenient. 
To further analyze sleep snoring, snoring events also need to be extracted from the sleep audio. 
However, the occurrence time of a snoring sound event is short and random, which presents a new challenge for sleep snoring-based OSAHS assessment.
\textbf{2) The speech signal is limited in reflecting upper airway obstruction information.}
The effectiveness of speech in diagnosing OSAHS is based on the fact that some speech articulation process involves parts of obstruction sites associated with OSAHS~\cite{ding2021severity}.
Meanwhile, the articulatory of snoring is by the flutter of OSAHS-related obstruction sites in the upper airway~\cite{quinn1995observation}.
This makes upper airway information contained in snores more applicable to OSAHS evaluation than in speech signals.

To tackle these problems, we present SimuSOE: a Simulated Snoring dataset for OSAHS Evaluation.
Simulated snoring is employed for OSAHS screening, which \textbf{1) can be effectively captured during wakefulness} and \textbf{2) may comprehensively reveal the pathological features of the upper airway}.
Participants simulate snoring several times by breathing deeply enough to vibrate the obstruction sites in their upper airway~\cite{mikami2012classification}. 
Compared with sleep snoring, simulated snoring can be recorded easily during wakefulness.
Previous works have demonstrated that simulated snoring has little difference in assessing the degree of obstruction~\cite{herzog2006prognostic,herzog2008frequency}, and it has already been applied in some medical research for OSAHS~\cite{ibrahim2014comparison,herzog2015drug}.
Compared with speech signals, simulated snoring may be more suitable for the OSAHS screening task.
It may provide a more responsive reflection of upper airway conditions during sleep, given the acoustic distinctions between speech and snoring~\cite{pevernagie2010acoustics}.
For patients with position-dependent OSAHS, sleep position could influence the site, direction, and severity of upper airway obstruction~\cite{vonk2019drug,carrasco2019drug}. Therefore, SimuSOE collects snoring data in both supine and lateral positions to improve the applicability of screening methods. 
To summarize, we make the following contributions:
\begin{enumerate}
    \item SimuSOE includes 4428 simulated snoring records from 82 adult participants, providing a time-efficient self-evaluation for OSAHS and contributing to expanded screening.
    \item We recorded simulated snoring in supine and lateral positions, enhancing OSAHS evaluation by comprehensively capturing upper airway obstruction.
    \item Our baseline experiments show that using simulated snoring is effective for OSAHS evaluation, and incorporating sleeping position information enhances screening.
\end{enumerate}



\begin{table*}[t]
\tiny
\caption{Physiological characterization of participants grouped by AHI classification thresholds. Using AHI = 5 event/h as a threshold, this group was used to examine the performance of simulated snoring on the OSAHS diagnosis task. AHI = 30 event/h as a threshold was applied to examine the performance of simulated snoring on the severe OSAHS screening task. The age, BMI, and AHI are present as mean $\pm$ standard deviation.}
\label{tab:data-distribution}
\resizebox{\textwidth}{!}{
\begin{tabular}{lcccccc}
\toprule
Threshold & Group                                                            & Male & Female & Age (years)     & BMI (kg/m$\tiny\rm{^2}$) & AHI (events/h)  \\ \midrule
\multirow{2}{*}{AHI = 5 events/h}  & AHI $\leq$ 5  & 7  & 6  & 24.85$\pm$4.49 & 21.78$\pm$1.73 & 1.52$\pm$1.01  \\
          & AHI $\textgreater$ 5  & 56   & 13     & 37.23$\pm$11.62 & 25.78$\pm$4.00     & 47.43$\pm$28.84 \\  \cmidrule{2-7}
\multirow{2}{*}{AHI = 30 events/h} & AHI $\leq$ 30 & 27 & 12 & 28.85$\pm$9.96 & 22.86$\pm$2.56 & 11.72$\pm$9.26 \\
          & AHI $\textgreater$ 30 & 36   & 7      & 41.09$\pm$10.02 & 27.21$\pm$3.97     & 65.94$\pm$19.89 \\  \bottomrule 
\end{tabular}
}
\end{table*}


\section{Datasets}
To construct the SimuSOE dataset, we process it in two steps. Firstly, we record simulated snoring signals in both supine and lateral positions for each subject. Subsequently, patients undergo PSG diagnosis, and experienced experts who get RPSGT\footnote{https://www.brpt.org/rpsgt/} (Registered Polysomnographic Technologists) certified generate PSG reports based on the recorded PSG data.

\subsection{Data Overview}
This study was conducted with the approval of the local medical ethics committee and adheres to the tenets of the Declaration of Helsinki. Informed consent was obtained from each participant. Personal information was collected and stored anonymously to ensure privacy protection.

We collect simulated snores and PSG reports data from 82 adult participants, including 63 males and 19 females, who undergo PSG diagnosis at a Sleep Medicine Centre within a local hospital. Patients with partial absence of PSG channel signals are excluded. 
SimuSOE comprises 4428 simulated snoring audio samples, with each snoring sound having a duration ranging from 0.55 to 9.84 seconds. The descriptive characteristics of the participants are shown in Table~\ref{tab:data-distribution}.

\subsection{Data Acquisition}
In this paper, we recorded simulated snoring with a subminiature lavalier microphone (Audio-Technica AT831R) driven by an audio Interface (Black Lion Audio Revolution 2x2). A sampling frequency of 44.1 kHz and a sampling resolution of 16 bits (bit depth) were used. The microphone, positioned 3 cm from the mouth, was directed towards the patient's mouth to ensure optimal recording conditions. This close-up recording enables it to capture subtle nuances in the snoring sounds and reduces background noise, thereby enhancing audio quality. Figure~\ref{fi:record} illustrates the data generation process.

Participants were asked to simulate snoring several times by breathing deeply enough to vibrate the soft palate, tonsils, tongue base, and epiglottis, which are the locations of upper airway obstruction in patients with OSAHS~\cite{janott2018snoring}. This recording procedure was performed in the supine, left lateral, and right lateral positions. 
In lateral positions, the microphone was repositioned to the opposite side of the face to avoid covering, ensuring the recording of a cleaner sound.

After simulated snoring sounds were recorded, each participant underwent a standard full PSG diagnosis. Following the guidelines of the AASM Manual V2.6, three experienced experts who got RPSGT certified, scored sleep stages and respiratory events based on the recorded PSG data. The resulting report included the AHI and other relevant physiological information. Patients were diagnosed and their severity was determined based on AHI. 
The physiological information includes age, gender, and Body Mass Index (BMI), among other details.

Based on the diagnostic results provided by the PSG report, we annotated the SimuSOE data. All simulated snores for each participant shared a common AHI label. Additionally, corresponding physiological information, such as age, gender, BMI, among other details, was annotated for future research purposes.



\subsection{Data Distribution}
The gender ratio of male to female in the SimuSOE dataset is approximately 3:1, ranges of BMI and age of all participants are $\rm 25.15 \pm 4.01\ kg/m^2$ and $\rm 35.27 \pm 11.71\ years$, respectively, as shown in Table~\ref{tab:data-distribution}. This data distribution aligns with the fact that OSAHS is more prevalent among males than females~\cite{daltro2006obstructive}, and since age and obesity are high-risk factors for OSAHS, older individuals with higher BMI are more likely to exhibit greater severity.

\begin{table}[t]
\caption{Description of the groups in each set of SimuSOE, grouped by AHI classification thresholds. Thresholds are established according to the OSAHS severity definitions outlined in the AASM Manual V2.6~\cite{berry2020aasm}.}
\label{tab:datasplit}
\resizebox{\columnwidth}{!}{%
\begin{tabular}{clcccc}
\toprule
                                       & \multicolumn{3}{l}{AHI = 5 events/h}                        & \multicolumn{2}{l}{AHI = 30 event/h} \\ \cmidrule(lr){2-4}  \cmidrule(lr){5-6}  
                                       & Description         & AHI $\leq$ 5   & AHI $\textgreater$ 5 & AHI $\leq$ 30 & AHI $\textgreater$ 30 \\ \midrule
\multirow{4}{*}{\rotatebox{90}{Train}} & Male                & 4              & 51                   & 22            & 33                   \\
                                       & Female              & 3              & 8                    & 7             & 4                    \\
                                       & avgAHI (events/h)   & 1.27           & 46.13                & 12.82         & 63.76                \\
                                       & Samples             & 378            & 3186                 & 1566          & 1998                 \\ \cmidrule(l){2-6} 
\multirow{4}{*}{\rotatebox{90}{Test}}  & Male                & 3              & 5                    & 5             & 3                    \\
                                       & Female              & 3              & 5                    & 5             & 3                    \\
                                       & avgAHI (events/h)   & 1.80           & 55.10                & 8.54          & 79.40                \\
                                       & Samples             & 324            & 540                  & 540           & 324                  \\ \bottomrule
\end{tabular}%
}
\end{table}

\section{Experiments}
The SimuSOE dataset is introduced to offer an efficient and cost-effective approach for the preliminary screening of OSAHS. Therefore, to assess the effectiveness of simulated snoring and sleeping positions, we perform two distinct OSAHS screening tasks using varied position sets.

\subsection{Experimental Setup}

\subsubsection{Data Split}
Inspired by Ding et al.~\cite{ding2021severity}, similar thresholds AHI = 5 events/h and AHI = 30 events/h were used in this study to separate the participants into groups.
As defined in AASM Manual V2.6, OSAHS or severe OSAHS was diagnosed when AHI $\textgreater$ 5 or AHI $\textgreater$ 30. Hence, we categorized the data into an OSAHS diagnostic group and a severe OSAHS screening group, determined by the AHI threshold.

The SimuSOE dataset contains a total of 4428 snoring samples from 82 participants. The training and test sets were divided based on the objectives of OSAHS screening and further assessing the effectiveness of different sleeping positions, as shown in Table~\ref{tab:datasplit}. 
For brevity, we utilized the same training and test set for AHI thresholds of 5 and 30, comprising 66 and 16 participants, respectively. 
For the test set, given the dataset imbalance with only 13 participants having AHI $\leq$ 5, we first randomly selected 6 OSAHS-negative subjects. Correspondingly, six participants with AHI $\textgreater$ 30 were then randomly chosen for the test set. To avoid extreme representation, where only OSAHS-negative subjects and severe patients are included in the test set, we also randomly selected four samples from those with AHI between 5 and 30 for the test set. 
For the training set, the remaining participants were allocated, comprising 66 subjects.

\begin{table}[t]
\caption{Classification results for participants grouped by AHI thresholds. The accuracy, sensitivity ($S_e$), specificity ($S_p$) and Score are present as mean $\pm$ standard deviation.}
\label{tab:result-lying}
\resizebox{\columnwidth}{!}{%
\begin{tabular}{lcccc}
\toprule
Threshold & \textbf{Accuracy} (\%) & $S_e$ (\%) & $S_p$ (\%) & Score (\%)     \\ \midrule
AHI = 5   & $\rm {67.36}_{\pm 1.77}$ & $\rm {96.85}_{\pm 0.94}$  & $\rm {18.21}_{\pm 6.11}$  & $\rm {57.53}_{\pm 2.63}$ \\
AHI = 30  & $\rm {70.49}_{\pm 1.24}$ & $\rm {62.04}_{\pm 6.59}$  & $\rm {75.56}_{\pm 1.98}$  & $\rm {68.80}_{\pm 2.31}$ \\  \bottomrule
\end{tabular}%
}
\end{table}

\subsubsection{Evaluation Metrics}
To demonstrate the ability of the SimuSOE dataset to screen OSAHS and severe OSAHS, we use \textbf{Accuracy} as the main evaluation metric for method performance. Following previous studies that use snores or speech signals to diagnose OSAHS~\cite{ding2021severity,xie2023assessment}, we also employ sensitivity ($S_e$), specificity ($S_p$), and Score as evaluation metrics. 
The score is defined as the average of $S_e$ and $S_p$, which reflects the performance of the method in diagnosing whether a participant is OSAHS/severe OSAHS positive.

\subsubsection{Training Details}
To assess the validity of the SimuSOE dataset, we employed the AST model as the baseline model~\cite{gong2021ast}. The simulated snoring sounds were recorded at 44.1 kHz and downsampling to 32 kHz. Before input to the AST encoder, we extracted 128-dimensional log Mel spectrograms for the input snore, using a Hamming window of 25 ms and a hop length of 10 ms.
After extracting the log Mel spectrograms, the simulated snoring recorded in different body positions by the same participant, one for each position, was selected and concatenated. A pooling layer was then utilized to aggregate the sequence of representations into a single vector that fits the input shape of AST.

The model was trained on one NVIDIA GeForce RTX 3090 GPU with batch size 8 for 30 epochs. The experimental setup was kept mostly consistent with the AST. We utilized binary cross-entropy loss and the Adam optimizer~\cite{kingma2014adam} with an initial learning rate of $\rm{1\times10^{-5}}$. The frequency/time mask data augmentation was used to train the model with a max time mask length of 48 frames and a max frequency mask length of 48 bins. Note that we present the results of our experiments over three random runs.

\begin{figure*}[t]   
  \centering           
  \subfloat[supine position]
  {
      \label{fig:cf_supine}\includegraphics[width=0.3\textwidth]{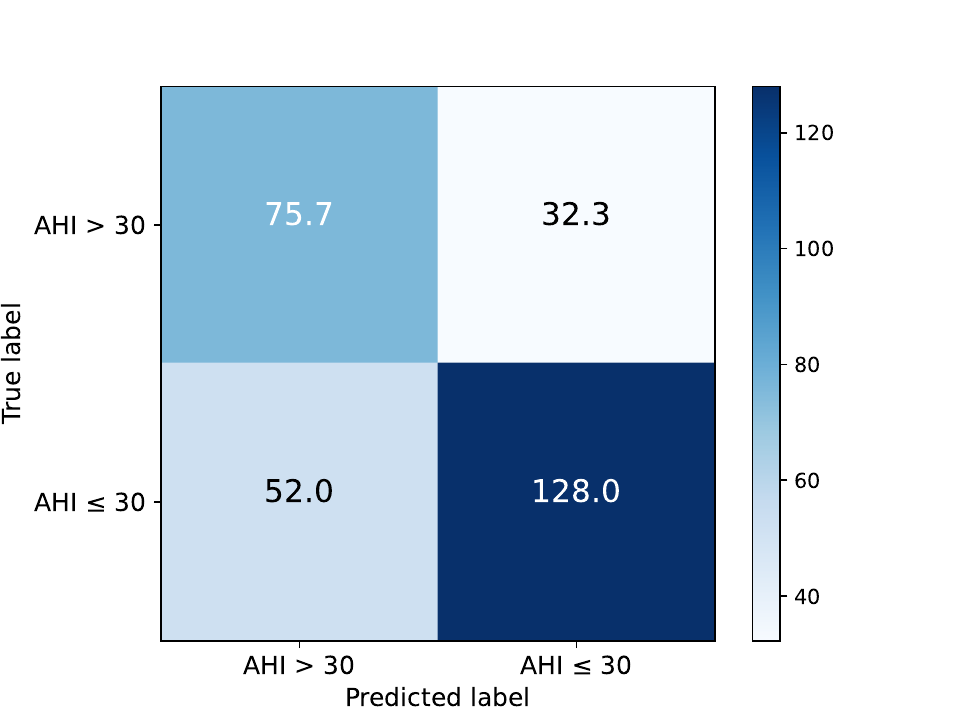}
  }
  \subfloat[lateral position]
  {
      \label{fig:cf_lateral}\includegraphics[width=0.3\textwidth]{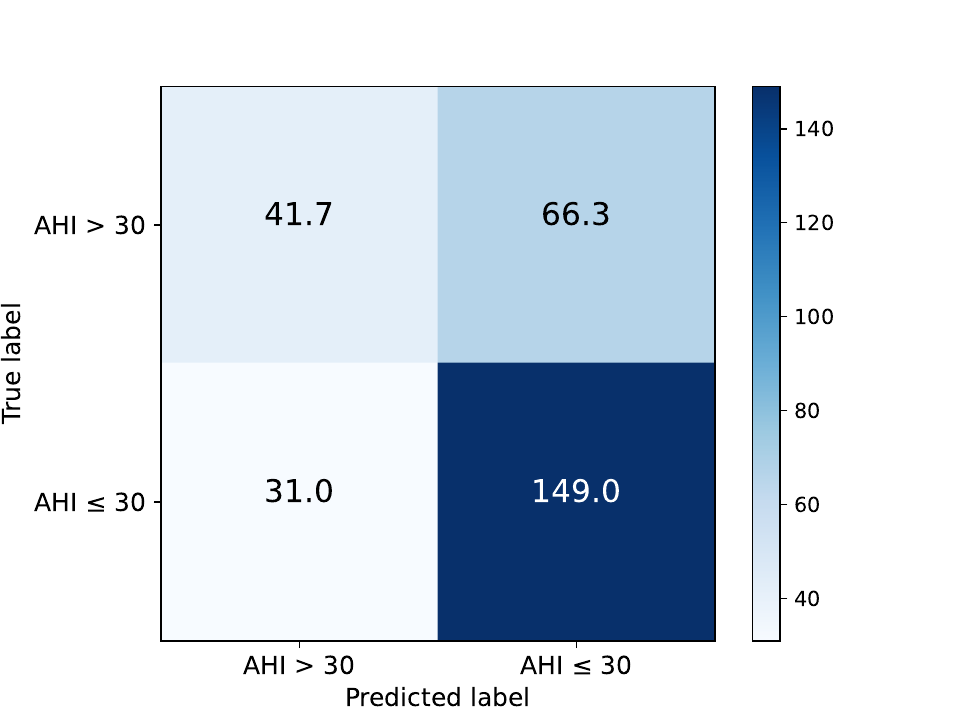}
  }
  \subfloat[lying position]
  {
      \label{fig:cf_lying}\includegraphics[width=0.3\textwidth]{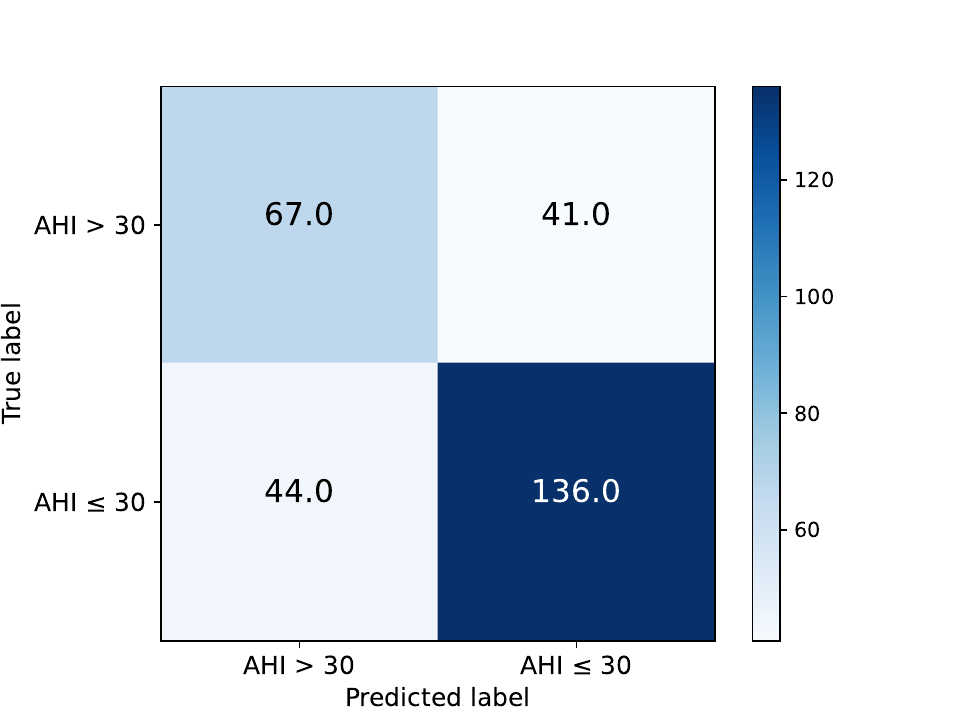}
  }
  \caption{Confusion matrix comparison of simulated snoring in different sleeping positions on the severe OSAHS screening task (threshold AHI = 30 events/h). Note that we present the confusion matrix of our experiments over three random runs.}
  \label{fig:cf} 
\end{figure*}

\subsection{Results}

\subsubsection{The classification effects of different screening tasks}
To evaluate the ability of simulated snoring in OSAHS diagnosis and severe OSAHS screening, we conduct two binary-class classification experiments on the SimuSOE dataset, with AHI thresholds set at 5 events/h and 30 events/h.

As shown in Table~\ref{tab:result-lying}, the model demonstrated an accuracy of 67.36\% and 70.49\% for group AHI = 5 events/h and group AHI = 30 events/h, respectively. This result indicates the utility of simulated snoring for diagnosing OSAHS and screening severe patients.
In the AHI = 5 events/h threshold group, the $S_e$ is notably elevated at 96.85\%, and a heightened $S_e$ is more desirable for medical classification tasks. Insufficient $S_e$ could result in missed detections, which potentially delay the treatment of patients and lead to more severe consequences.
However, the $S_p$ was 18.21\%, which is suboptimal. This discrepancy could be attributed to the fact that the AHI $\leq$ 5 group only has 13 subjects, which is significantly less than 69 subjects in the AHI $\textgreater$ 5 group. The imbalanced distribution of data poses a challenge for the model in effectively learning the upper airway characteristics of OSAHS-negative participants.
In the AHI = 30 events/h threshold group, the data within the group were more balanced, with a ratio close to 1:1, thus achieving a higher accuracy and Score at 70.49\% and 68.80\%. Consequently, the experimental results in this group demonstrated a more balanced trade-off between $S_e$ and $S_p$, indicating that the model is more competent at screening severe OSAHS subjects.

\begin{table}[t]
\caption{Performance comparison of simulated snoring in different sleeping positions on the OSAHS classification task. The results in the lying position group were obtained by learning simulated snoring in both supine and lateral positions. \textbf{Best} and \underline{second best} results.}
\label{tab:result-position}
\resizebox{\columnwidth}{!}{%
\begin{tabular}{lccccc}
\toprule
Threshold                 & Position         & \textbf{Accuracy} (\%)   & $S_e$ (\%)                & $S_p$ (\%)                & Score (\%)     \\ \midrule
\multirow{3}{*}{AHI = 5}  & supine           & $\rm {64.70}_{\pm 0.59}$ & $\rm {97.22}_{\pm 0.79}$  & $\rm {10.49}_{\pm 2.86}$  & $\rm {53.86}_{\pm 1.04}$ \\
                          & lateral          & $\rm {\underline{65.05}}_{\pm 2.87}$ & $\rm {97.59}_{\pm 2.05}$  & $\rm {10.80}_{\pm 5.03}$  & $\rm {\underline{54.20}}_{\pm 3.26}$ \\
                          & lying & $\rm {\textbf{67.36}}_{\pm 1.77}$ & $\rm {96.85}_{\pm 0.94}$  & $\rm {18.21}_{\pm 6.11}$  & $\rm {\textbf{57.53}}_{\pm 2.63}$ \\ \cmidrule{2-6}
\multirow{3}{*}{AHI = 30} & supine           & $\rm {\textbf{70.72}}_{\pm 0.43}$ & $\rm {70.06}_{\pm 6.06}$  & $\rm {71.11}_{\pm 3.27}$  & $\rm {\textbf{70.59}}_{\pm 1.45}$ \\
                          & lateral          & $\rm {66.20}_{\pm 2.13}$ & $\rm {38.58}_{\pm 11.37}$ & $\rm {82.78}_{\pm 6.30}$  & $\rm {60.68}_{\pm 3.34}$ \\
                          & lying & $\rm {\underline{70.49}}_{\pm 1.24}$ & $\rm {62.04}_{\pm 6.59}$  & $\rm {75.56}_{\pm 1.98}$  & $\rm {\underline{68.80}}_{\pm 2.31}$ \\ \bottomrule
\end{tabular}%
}
\end{table}

\subsubsection{Classification effects in different sleeping positions}

To delve deeper into the impact of sleep body position on the OSAHS screening task, we divided the data based on sleeping position.
This experiment aims to validate the hypothesis that sleep body position is a significant factor in the OSAHS screening task, and upper airway obstruction features extracted from simulated snoring in the lateral position could serve as a complementary source of information.
Based on AHI groups, we further segregated the data into the supine group and the lateral group, and we compared the results of these two groups with those combining both sleeping positions. 

The experimental results for both OSAHS diagnosis and severe subjects screening task are presented in Table~\ref{tab:result-position}. 
In the OSAHS diagnosis group, the accuracies of the supine, lateral, and lying (combined position) groups were 64.70\%, 65.05\%, and 67.36\%, respectively. Compared to the supine position group, the classification accuracy in the lying position group increased by 2.66\%. This enhancement confirms that an important relationship exists between sleep body position and the snoring sounds generated~\cite{xiao2023snoring}. Furthermore, this result also demonstrates that the airway obstruction features extracted from simulated snoring in the lateral position can serve as a complementary factor in OSAHS diagnosis.
  
However, in the severe OSAHS screening group, the accuracy of the lying position group is slightly lower than that of the supine group by 0.23\%. We speculate that this is mainly attributed to the fact that most OSAHS-related snoring events occur in the supine position~\cite{carrasco2019drug}, especially prevalent among patients with severe OSAHS. While simulated snoring in the lateral position contributes additional information to the OSAHS diagnosis task, it introduces some interference information in the severe OSAHS screening task.
As illustrated in Figure~\ref{fig:cf_lateral}, simulated snoring in the lateral position shows better performance in identifying non-severe OSAHS subjects (AHI $\leq$ 30), while its effectiveness is not as prominent in screening severe OSAHS subjects. This pattern is consistent with the recording process, where severe patients were more prone to generate simulated snoring when in the supine position, in contrast to other participants, particularly OSAHS-negative subjects, who did not exhibit such a tendency.

\section{Conclusions}
This study presented an OSAHS evaluation dataset referred to as SimuSOE, comprising simulated snoring records in different sleep body positions.
By introducing simulated snoring as an alternative to sleep snoring, the aim was to provide a precise and rapid method for the initial screening of OSAHS.
The OSAHS evaluation result on the SimuSOE dataset achieved an accuracy of 70.72\% with a score of 70.59\%, indicating that the analysis of simulated snoring during wakefulness can serve as an effective preliminary screening method for OSAHS. Moreover, experimental results demonstrated that analyzing simulated snoring in both supine and lateral positions contributes to enhancing OSAHS diagnosis accuracy. 
Further progress can be anticipated by augmenting participants within this dataset to alleviate the impact of data imbalance and combining multiple physiological signals to develop new methods for modeling snoring classification.
Our dataset will be available when the applicant submits a formal document.

\noindent \textbf{Acknowledge.}
This work was supported in part by the National Nature Science Foundation of China (No.62071342, No.62171326) and the Hubei Province Technological Innovation Major Project (No.2022BCA041).
\bibliographystyle{IEEEtran}
\bibliography{mybib}

\end{document}